\documentclass[aps,prl,reprint,groupedaddress]{revtex4-1}
\usepackage{graphicx}% Include figure files
\usepackage{dcolumn}% Align table columns on decimal point
\usepackage{bm}% bold math
\usepackage{color}
\usepackage{array}
\usepackage{mathrsfs}
\usepackage{ulem} % sout
\usepackage{amsmath}
\usepackage{amssymb}
\usepackage{url}

\begin{document}

%\preprint{}

\title{Ground-state phase diagram of the Kitaev-Heisenberg model on a kagome lattice}



%\thanks{}

\author{Katsuhiro Morita}
\email[e-mail:]{katsuhiro.morita@rs.tus.ac.jp}
\affiliation{Department of Applied Physics, Tokyo University of Science, Tokyo 125-8585, Japan}

\author{Masanori Kishimoto}
\affiliation{Department of Applied Physics, Tokyo University of Science, Tokyo 125-8585, Japan}

\author{Takami Tohyama}
\affiliation{Department of Applied Physics, Tokyo University of Science, Tokyo 125-8585, Japan}


\date{\today}

\begin{abstract}
The Kitaev-Heisenberg model on the honeycomb lattice has been studied for the purpose of finding exotic states such as quantum spin liquid and topological orders.
On the kagome lattice, in spite of a spin-liquid ground state in the Heisenberg model, the stability of the spin-liquid state has hardly been studied in the presence of the Kitaev interaction. Therefore, we investigate the ground state of the classical and quantum spin systems of the kagome Kitaev-Heisenberg model. 
In the classical system, we obtain an exact phase diagram that has an eight-fold degenerated canted ferromagnetic phase and a subextensive degenerated Kitaev antiferromagnetic phase. 
In the quantum system, using the Lanczos-type exact diagnalization and cluster mean-field methods,  we obtain two quantum spin-liquid phases, an eight-fold degenerated canted ferromagnetic phase similar to the classical spin system, and an eight-fold degenerated $\bf q=0$ $120^\circ$ ordered phase induced by quantum fluctuation. 
These results may provide a crucial clue to recently observed magnetic structures of the rare-earth-based kagome lattice compounds $A_2$RE$_3$Sb$_3$O$_{14}$ ($A$ = Mg, Zn; RE = Pr, Nd, Gd, Tb, Dy, Ho, Er, Yb).
\end{abstract}
%\keywords{}

\maketitle
The Kitaev-Heisenberg (KH) model and related models on the various lattices have been studied theoretically for the sake of realization of exotic states such as quantum spin liquid and topological orders~\cite{kitaev2006,rousochatzakis2015,lee2014,nasu2014,takayama2015,yao2007,kargarian2010,jahromi2016,jahromi2018,kishimoto2018,kimchi2014}. 
The models have also been examined in connection with real compounds such as iridium oxides with honeycomb and triangular lattices~\cite{kimchi2014,chaloupka2010,jiang2011,reuther2011,kimchi2011,schaffer2013,singh2012,choi2012,reuther2014,bhattacharjee2012,chaloupka2013,okamoto2013,price2013,sela2014,katukuri2014,rau2014,rau2014b,yamaji2014,
sizyuk2014,kimchi2015,shinjo2015,chaloupka2015,okubo2017,gotfryd2017,dey2012,li2015,becker2015,catuneanu2015,rousochatzakis2016,shinjo2016,lee2017,haraguchi2018} 
and $\alpha$-RuCl$_3$~\cite{plumb2014,kubota2015,kim2015,johnson2015,bastien2018,yadav2016,zhu2018,gohlke2018}.

Concerning spin-liquid phase, the spin-1/2 Heisenberg  model on the kagome lattice (KL) is another candidate. The model has been studied for several decades~\cite{waldtmann1998,jiang2008,yan2011,depenbrock2012,nishimoto2013,mei2017,ran2007,iqbal2011,he2017,liao2017,helton2007,olariu2008,helton2010,han2012,feng2017,morita2008,ono2009,matan2010,ono2014,goto2016,okuma2017}, and it has theoretically been predicted that the ground state becomes the gapped 
$\mathbb{Z}_2$ spin liquid~\cite{yan2011,depenbrock2012,nishimoto2013,mei2017} or gapless U(1) spin liquid~\cite{ran2007,iqbal2011,he2017,liao2017}. 
Moreover, the kagome spin models with strong anisotropy in exchange interactions, such as the kagome ice model, have been studied as well~\cite{damle2006,qi2008,sendetskyi2016,benjamin2016,changlani2018,seshadri2018}.
In spite of such intensive studies on the kagome Heisenberg model, the effect of the Kitaev interaction on the spin-liquid state has hardly been studied~\cite{kimchi2014}. Furthermore, the precise phase diagram is not presented.

\begin{figure}[tb]
  \begin{center}
	\includegraphics[width=80mm]{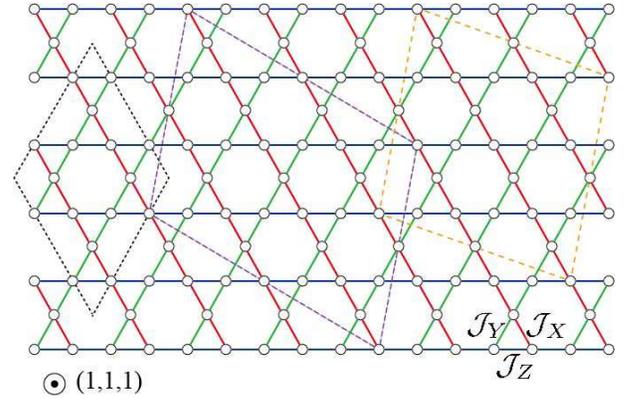}
    \caption{Lattice structure of the KL with three anisotropic exchange interactions, $\mathcal{J}_X$, $\mathcal{J}_Y$, and $\mathcal{J}_Z$. The red, green, and blue solid lines denote $\mathcal{J}_X$, $\mathcal{J}_Y$, and $\mathcal{J}_Z$, respectively. The black, orange, and purple dashed quadrangles denote the clusters of $N=12$, 24 and 30, respectively, used in ED method with periodic boundary conditions.}
    \label{model}
  \end{center}
\end{figure}

Recently, the rare-earth-based KL compounds $A_2$RE$_3$Sb$_3$O$_{14}$ ($A$ = Mg, Zn; RE = Pr, Nd, Gd, Tb, Dy, Ho, Er, Yb) whose space group is $R\overline{3}m$
 have been synthesized~\cite{dun2016,scheie2016,paddison2016,sanders2016,sanders2016b,dun2017}.
The compounds except for RE = Gd have an effective spin with $S=1/2$ on the KL because of the Kramers or non-Kramers doublet ground state~\cite{dun2017}. The exchange interactions between the nearest-neighbor (NN) spins are expected to be anisotropic~\cite{dun2017}, that is, magnetic interaction between the NN sites depends on the bond of a triangular unit in the lattice (see Fig~\ref{model}).
Although these compounds would not have Kitaev-type interactions, the effective Hamiltonian of these compounds must contain the symmetry of $R\overline{3}m$.
Since the kagome KH model contains this symmetry together with bond-dependent anisotropic interactions, 
there is a possibility that the phase obtained in the KH model are continuously connected with that in the effective models of the compounds.
Therefore, the study of this model will contribute to the understanding of $A_2$RE$_3$Sb$_3$O$_{14}$. %though there is no report on the presence of Kitaev-type interactions at present.
Hence, it is necessary to perform a theoretical investigation in order to elucidate its ground-state properties and to find new novel phases.

In this paper, we investigate the ground state of the classical and quantum spin systems of the kagome KH model.
In the classical spin system, we obtain an exact phase diagram based on the analytical solution of a three-spin cluster. 
We confirm two kinds of phases that are an eight-fold degenerated canted ferromagnetic (CFM) and a subextensive degenerated Kitaev antiferromagnetic (KAF).
In between CFM and KAF, there are the Heisenberg point and its Klein duality point whose ground state is a classical spin liquid (CSL) state.
For the quantum spin system, we use the Lanczos-type exact diagnalization (ED) and cluster mean-field (CMF) methods.
We find that the spin-liquid state in the kagome Heisenberg model remains even for small Kitaev-type interaction. We also find an eight-fold degenerated CFM phase similar to the classical spin system, and an eight-fold degenerated $\bf q=0$ $120^\circ$ ordered phase induced by quantum fluctuations, which corresponds to the subextensive degenerated KAF in the classical spin system. It is interesting that $A_2$RE$_3$Sb$_3$O$_{14}$ ($A$ = Mg, RE = Gd, Er) and ($A$ = Mg, RE = Nd) have the same type of the $\bf q=0$ $120^\circ$ order~\cite{dun2016} and the CMF~\cite{scheie2016}, respectively.

The Hamiltonian of the KH model on the KL is given by
\begin{equation}
  \mathcal{H}= \sum _{\langle i,j \rangle} {\bf S}_i ^{\rm T}  \mathcal{J}_{i,j}{\bf S}_j \label{hamiltonian},
\end{equation}
where ${\bf S}_i$ is a classical spin vector ${\bf S}_i=(S_i^x\  S_i^y\ S_i^z)^{\rm T} \in \mathbb{R} ^3$ with $|{\bf S}_i|=1$ (a quantum spin operator with $S=1/2$) at site $i$ for classical (quantum) system. $\mathcal{J}_{i,j}$ represents the NN interactions as shown in Fig~\ref{model} and takes one of the three anisotropic interactions, $\mathcal{J}_X=\mathrm{diag}(J+K, J, J)$, $\mathcal{J}_Y = \mathrm{diag}(J, J+K, J)$, and $\mathcal{J}_Z=\mathrm{diag}(J, J, J+K)$, where $K$ and $J$ correspond to the energy of the Kitaev and Heisenberg interactions, respectively.  We note that there is the Klein duality~\cite{kimchi2014} in this model, which transforms $(J,K) \mapsto ( \tilde{J}, \tilde{K})=(-J, 2J+K)$. 
We introduce the parametrization $(J,K) = (I\cos\theta,I\sin\theta)$, where $I$ is the energy unit ($I=1$).

\begin{figure}[tb]
  \begin{center}
\includegraphics[width=80mm]{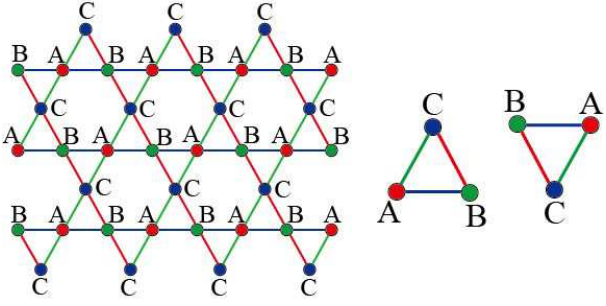}
    \caption{Three-sublattice pattern of the KL (left).  Two triangles (right) are equivalent to all triangles in the KL. The red circle $A$, green circle $B$, and blue circle $C$ represent independent sublattice. The red, green, and blue solid lines denote $\mathcal{J}_X$, $\mathcal{J}_Y$, and $\mathcal{J}_Z$, respectively.}
    \label{3sub}
  \end{center}
\end{figure}

We first determine the exact classical phase diagram.
The KL can be divided into three sublattices (three color sites with classical spins, ${\bf S}_A$, ${\bf S}_B$, and ${\bf S}_C$, forming a triangle) as shown in Fig~\ref{3sub}.
All of the triangles in the KL have the same structure as two triangles shown in the right-hand side of Fig~\ref{3sub}. 
The KH Hamiltonian on a triangle is given by
\begin{equation}
  h_{\Delta} =  {\bf S}_A ^{\rm T}  \mathcal{J}_{Z}{\bf S}_B +  {\bf S}_B ^{\rm T}  \mathcal{J}_{X}{\bf S}_C  +   {\bf S}_C ^{\rm T}  \mathcal{J}_{Y}{\bf S}_A \label{trihami}.
\end{equation}
Then, the Hamiltonian (\ref{hamiltonian}) reads $\mathcal{H}= \sum _{\Delta} h_{\Delta}$,
where the summation is performed for all triangles (not only upward triangles but also downward triangles in Fig.~\ref{3sub}).
A special solution for the ground state of $\mathcal{H}$ can be obtained by covering the KL with the ground-state vectors (${\bf S}_A$, ${\bf S}_B$, ${\bf S}_C$) of $h_\Delta$~(\ref{trihami}), because all triangles on the KL have the lowest energy.
The ground-state vectors (${\bf S}_A$, ${\bf S}_B$, ${\bf S}_C$) and energy $E_{\rm min{\Delta}}$  are given as follows:
in $\theta \in [0,\pi -\arctan(2)]$,
${\bf S}_A = (0,\frac{c_y}{\sqrt{2}},\frac{c_z}{\sqrt{2}})$,   
${\bf S}_B = (\frac{c_x}{\sqrt{2}},0,-\frac{c_z}{\sqrt{2}})$,   
${\bf S}_C = (-\frac{c_x}{\sqrt{2}},-\frac{c_y}{\sqrt{2}},0)$, and 
$E_{\rm min{\Delta}} = -\frac{3}{2} \left( \sin \theta + \cos \theta \right)$, while
in $\theta \in [\pi -\arctan(2),2\pi]$, 
${\bf S}_A = \left(c_xF(\theta),c_yG(\theta),c_zG(\theta)\right)$, 
${\bf S}_B =  \left(c_xG(\theta),c_yF(\theta),c_zG(\theta)\right)$, 
${\bf S}_C =  \left(c_xG(\theta),c_yG(\theta),c_zF(\theta)\right)$, 
and $E_{\rm min{\Delta}} = \frac{3}{4} \left( \sin \theta + \cos \theta \right) - \frac{3}{4} \sqrt{ \sin2\theta + 4\cos2\theta +5 }$,
where $c_x,c_y,c_z \in \{ -1,1 \}$,
$F(\theta) = f(\theta)/\sqrt{f(\theta)^2+1}$ and
$G(\theta) = 1/\sqrt{f(\theta)^2+1}$ with 
$ f(\theta) = 4\cos\theta/(\cos\theta+\sin\theta -\sqrt{\sin2\theta+4\cos2\theta+5})$.
%$F(\theta) = \cos\left(\frac{3}{4}\pi+\frac{1}{2}\alpha(\theta) \right)$ and
%$G(\theta) = \frac{1}{\sqrt{2}}\sin\left(\frac{3}{4}\pi+\frac{1}{2}\alpha(\theta) \right)$ with 
%$\alpha(\theta) = {\rm arctan2}\left(\sqrt2(\cos\theta+\sin\theta),4\cos\theta\right)$.
We note that the wave function at the Heisenberg points, i.e.,  $\theta=0$ and $\theta=\pi$, and their dual points, i.e., $\theta=\pi-\arctan(2)$ and $\theta=-\arctan(2)$, has the global rotation symmetry. %%%%%%koko

\begin{figure}[tb]
  \begin{center}
\includegraphics[width=86mm]{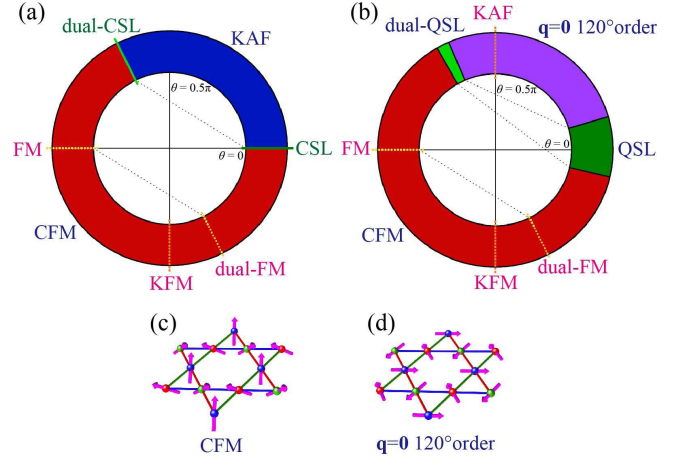}
    \caption{(a) Classical ground-state phase diagram of the KH model on the KL parametrized by $\theta$. The KAF and CFM phases are separated by the CSL (green line) and the dual one (light green line), while the FM point and its dual one (yellow dotted lines) as well as the KFM point (orange dotted line) do not change phases. 
(b) Quantum ground-state phase diagram obtained by the CMF method with $N=24$ cluster.
The CFM and $\bf q=0$ $120^\circ$ ordered phases are separated by the QSL phase and the dual one. The FM point and its dual one (yellow dotted lines) as well as the KAF and KFM points (orange dotted lines) do not change phases. In (a) and (b), the dotted lines in the circle connect dual points due to the Klein duality. 
The spin structure for the CFM and $\bf q=0$ $120^\circ$ ordered phases are schematically denoted in (c) and (d), respectively.}
    \label{PH}
  \end{center}
\end{figure}

Figure~\ref{PH}(a) shows  the exact classical ground-state phase diagram of the kagome KH model.
There are only two phases: one is an eight-fold degenerated CFM phase for $\theta \in [\pi -\arctan(2),2\pi]$ and the other is a $2^{3L}$-fold degenerated KAF phase for $\theta \in [0,\pi -\arctan(2)]$, where $L$ is the linear system size. 
This $2^{3L}$-fold degeneracy is caused by the absence of one of the three components in ${\bm S}_i$, which leads to $2^L$ degeneracy for each direction of the three bonds. In between the CFM and KAF phases, there is a macroscopic degenerated CSL state and a dual-CSL state corresponding to the Heisenberg and its dual points, respectively. There is no phase change across the Kitaev ferromagnetic (FM) point, i.e., $\theta=\frac{3}{2}\pi$, as well as the FM and its dual FM points. The Kitaev FM point gives a Kitaev ferromagnetic (KFM) state that  has $2^{3L}$-fold degeneracy as is the case of the KAF state.

\begin{figure}[tb]
  \begin{center}
\includegraphics[width=86mm]{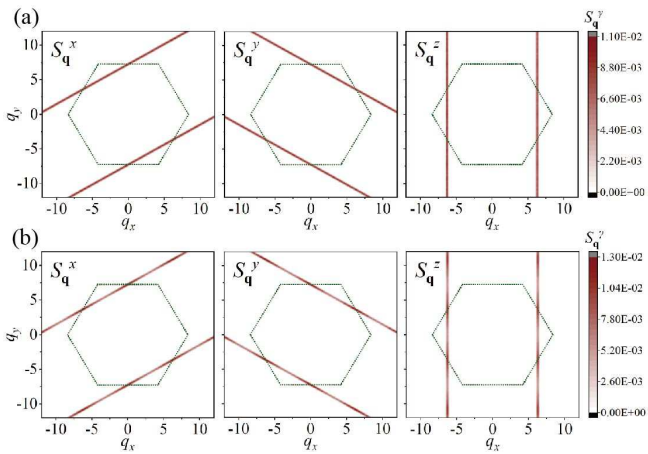}
    \caption{Static spin structure factor $\bf{S}_{\bf{q}}$ in the classical system at (a) $\theta=0.5\pi$ and (b) $\theta=0.3\pi$ in the KAF phase, obtained by the Monte Carlo simulation for an $N=24\times24\times3$ (=1728) lattice with periodic boundary conditions at $T=0.0005$. The green dotted hexagons denote the extended first Brillouin zone.}
    \label{Sq}
  \end{center}
\end{figure}

When the ground states are degenerate in the classical spin systems, thermal and quantum fluctuations can induce ordered states through so-called ``order-by-disorder'' mechanism.
Therefore, we calculate static spin structure factor $\bf{S_q}$ at finite temperature to confirm the effect of thermal fluctuations in the KAF phase using the Monte Carlo simulation.
All three components of $\bf{S_q}$ at $T=0.0005$ for $\theta=0.5\pi$ and $\theta=0.3\pi$ have linear distributions in the $\bf{q}$ space as shown in Fig.~\ref{Sq}.
This behavior is different from that of the triangular lattice where one of the three directions has stronger intensity because of the nematic order~\cite{rousochatzakis2016}.
The intensity of $\bf{S_q}$ at $\theta=0.5\pi$ is constant on the lines.
On the other hand, the intensity at $\theta=0.3\pi$ slightly increases when the lines cross the extended first Brillouin zone, which indicates a tendency toward the $\bf q=0$ $120^\circ$ order which will be discussed below.
Therefore, the change from $\theta=0.5\pi$ with only $K$ to $\theta=0.3\pi$ with finite $J$ indicates that the ``order-by-disorder'' in the KAF phase of the KL will be organized by $J$.

\begin{figure}[tb]
  \begin{center}
\includegraphics[width=86mm]{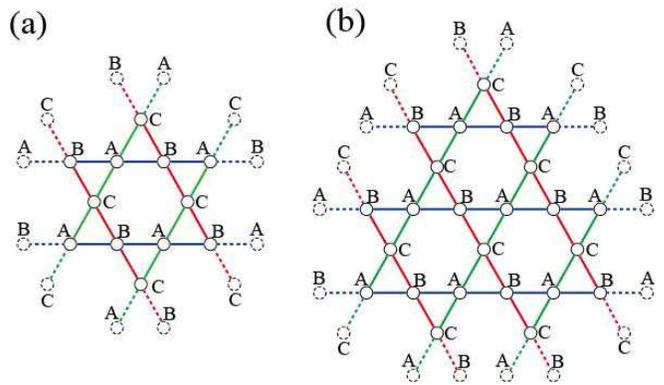}
    \caption{Three-sublattice structure with sublattices $A$, $B$, and $C$ used in our CMF method. (a) $N=12$ and (b) $N=24$. Dotted circles and dotted lines denote the mean field sites and mean field bonds, respectively.}
    \label{CMFlattice}
  \end{center}
\end{figure}

Next, we investigate the quantum system using the Lanczos-type ED and CMF methods. 
Our ED calculations are performed for the KL of $N= 12$, 24, and 30 with the periodic boundary conditions as shown in Fig.~\ref{model}, and our CMF method utilizes the $N= 12$ and 24 clusters as shown in Fig.~\ref{CMFlattice}.
We note that the CMF method has been successfully applied to the analysis of not only the Heisenberg model \cite{ren2014,yamamoto2015,yamamoto2016} but also the KH model \cite{gotfryd2017}.
We apply the standard mean-field approximation for interactions between the cluster-edge spin at site $i$, and the spin at a mean-field site belonging to a sublattice $\mu$: ${\bf{S}}_i\cdot{\bf{S}}_\mu \mapsto {\bf{S}}_i\cdot \langle \overline{\bf{S}_{\mu}} \rangle$ with $\langle \overline{\bf{S}_{\mu}} \rangle = N_{\mu}^{-1}\sum^{N_{\mu}}_{i_{\mu}}{ \langle {\bf{S}}_{i_\mu} \rangle}$,
where $\mu\in \{\rm{A,B,C}\}$ as shown in Fig.~\ref{CMFlattice}, and $i_{\mu}$ represents a site belonging to $\mu$ whose total number is $N_{\mu}$.
Thus, the CMF Hamiltonian reads
\begin{equation}
  \mathcal{H}_{\rm CMF}= \sum _{\langle i,j \rangle} {\bf S}_i ^{\rm T}  \mathcal{J}_{i,j}{\bf S}_j + \sum _{\langle i,{\mu} \rangle} {\bf S}_i ^{\rm T}  
\mathcal{J}_{i,{\mu}}\langle \overline{\bf{S}_{\mu}}\rangle, \label{CMFH}
\end{equation}
where the first and second terms represent intra-cluster interaction and the mean-field interaction, respectively.
$\langle \overline{\bf{S}_{\mu}}\rangle$ is self-consistently determined by applying the ED technique to $\mathcal{H}_{\rm CMF}$ (\ref{CMFH}).
The ground state is so obtained as to give the minimum CMF energy.
As shown in Fig.~\ref{CMFlattice}, we use a three-sublattice structure for both the $N= 12$ and 24 clusters.
We note that, even if a twelve-sublattice structure was used in the same clusters, the three-sublattice ordered state has the lowest energy. We also note that at the Kitaev points ($\theta=\pm0.5\pi$) the ground state has $2^{3L}$ degeneracy with $L=2$ in the twelve-sublattice structure.

\begin{figure}[tb]
  \begin{center}
\includegraphics[width=86mm]{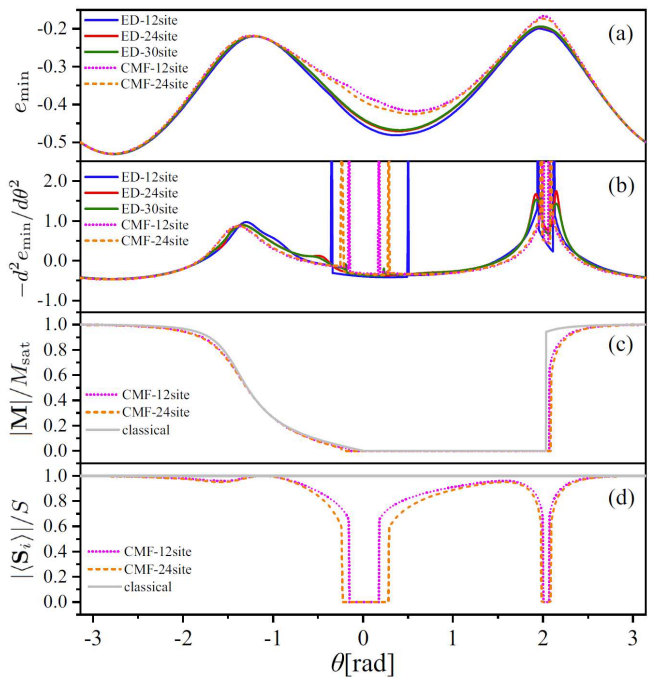}
    \caption{Quantum ground state energy per site, $e_{\rm min}$ (a), and its second derivative with respect to $\theta$, $-d^2 e_{\rm min}/d \theta^2$ (b), obtained by the ED and CMF methods. Normalized magnetization $|\bf{M}|/$$M_{\rm sat}$ (c) and  normalized local moment $|\langle \bf{S}$$_i\rangle|/S$ (d) for the quantum systems obtained by the CMF method and for the classical system.}
    \label{E-the}
  \end{center}
\end{figure}

The ground-state energies per site, $e_{\rm min}$, and its second derivatives with respect to $\theta$, $-d^2 e_{\rm min}/d \theta^2$, obtained by the ED and CMF methods are shown in Figs.~\ref{E-the}(a) and~\ref{E-the}(b).
There are two singularities in $d^2 e_{\rm min}/d \theta^2$ on either side of $\theta=0$ as well as on either side of its Klein duality point $\theta=\pi-\arctan(2)$ for the $N=12$ and 30 ED and $N=12$ and 24 CMF results.
These singularities come from the level crossing of the ground state.
In the ED results for $N=24$, there is no singularity but peak or hump structures are seen near the level crossing points.
Furthermore, the separation of the two singular points slightly increases with increasing the system size in the CMF results.
Therefore, these results indicate the presence of phase transitions on either side of $\theta=0$ and its duality point $\theta=\pi-\arctan(2)$.
If the level crossing remains in the thermodynamic limit, the phase transition will be the first-order one.
We note that the region near $\theta=0$ corresponds to a gapped quantum spin liquid (QSL) state 
in our calculation because of the presence of a finite gap between the ground state and the first excited state. 
If the quantum spin liquid (QSL) at $\theta=0$ has a finite energy gap in the thermodynamic limit \cite{yan2011,depenbrock2012,nishimoto2013,mei2017}, 
the gapped QSL state will remain near $\theta=0$ even in the presence of the Kitaev interaction. 

Figure~\ref{PH}(b) shows the phase diagram obtained by the $N=24$ CMF calculation. 
There is now a solid consensus of the presence of the spin-liquid ground state at the Heisenberg point ($\theta=0$)~\cite{yan2011,depenbrock2012,nishimoto2013,mei2017,ran2007,iqbal2011,he2017,liao2017}. 
Since there is no anomaly in $d^2 e_{\rm min}/d \theta^2$ across $\theta=0$ in Fig.~\ref{E-the}(b), the region including the Heisenberg point would be a QSL phase.
In this region, the normalized magnetization $|\bf{M}|/$$M_{\rm sat}$, where $M_{\rm sat}$ is the saturated magnetization,  and the normalized local moment $|\langle \bf{S}$$_i\rangle|/S$ are zero as shown in Figs.~\ref{E-the}(c) and~\ref{E-the}(d), as expected from QSL.
We note that, in the classical system, $|\langle \bf{S}$$_i\rangle|/S=1$, being independent of $\theta$. 
The corresponding dual region also shows the same zero value, indicating possible dual QSL state.

The region including the FM and dual FM points is an eight-fold degenerated CFM phase with wave vector $\bf q=0$ as is the case of the classical spin system.
This is evident from finite value of both $|\bf{M}|/$$M_{\rm sat}$ and $|\langle \bf{S}$$_i\rangle|/S$ in Figs.~\ref{E-the}(c) and~\ref{E-the}(d).
On the other hand, we find that the KAF phase in the classical system is replaced by an eight-fold degenerated $\bf q=0$ $120^\circ$ ordered phase due to quantum fluctuation. 
This order should have zero magnetization, which is clearly seen in Fig.~\ref{E-the}(c).
A tendency toward the order even in the classical system was discussed above as evidenced by the modulation of the intensity in $\bf{S_q}$ shown in Fig.~\ref{Sq}(b).

At the Kitaev points ($\theta=\pm0.5\pi$) in Fig.~\ref{PH}(b), there is neither tendency toward the CFM order nor toward the $\bf q=0$ $120^\circ$ order.
This indicates that quantum fluctuation induces the $\bf q=0$ orders only when Heisenberg interaction $J$ is present.
This tendency is also supported by the previous works~\cite{jiang2008,simokawa2016}.

Finally, we compare our results of the KH model on the KL with the experimental results of the rare-earth KL compounds $A_2$RE$_3$Sb$_3$O$_{14}$.
The CFM structure in $A$=Mg, RE = Nd and the $\bf q=0$ $120^\circ$ ordered structure in $A$=Mg, RE = Gd and Er have been observed by the neutron scattering experiments~\cite{dun2016,scheie2016}.  
We note that, although the CFM state can be obtained by introducing the Dzyaloshinsky-Moriya (DM) interactions,  
the $\bf q=0$ $120^\circ$ ordered structure in Fig.~\ref{PH}(d) cannot be realized even in the presence of the DM interactions~\cite{Elhajal2003}.
We thus speculate that the latter structure is continuously connected with the $\mathbf{q}=120^\circ$ order obtained in our KH model.
In $A$=Mg and RE =  Dy, there is a spin ice structure~\cite{dun2016,paddison2016}, which dose not exist in the KH model.
Therefore, we need to add further interactions, for example, so-called $\Gamma$ term~\cite{rau2014} and dipole interactions, to correctly analyze the compound.
 
In summary, inspired by the recent development of the KH model on various lattices and the discovery of the rare-earth-based KL compounds with anisotropic exchange interactions, we investigated the ground state of the classical and quantum ($S$=1/2) spin KH model on the KL.
In the classical system, we obtained the exact phase diagram with two kinds of phases that are the eight-fold degenerated CFM and the subextensive degenerated KAF.
In the quantum system, we found two QSL phases, the eight-fold degenerated CFM phase similar to the classical spin system, and the eight-fold degenerated $\bf q=0$ $120^\circ$  ordered phase induced by quantum fluctuations, which corresponds to the subextensive degenerated KAF in the classical spin system.
Moreover, we confirmed that the QSL state expected at the Heisenberg limit remains even in the presence of the small Kitaev interaction.
$A_2$RE$_3$Sb$_3$O$_{14}$ for $A$ = Mg and RE = Gd, Er ($A$ = Mg and RE = Nd) have the $\bf q=0$ $120^\circ$ order (the CMF) similar to our results. %, we can expect that these compounds may be similar to the KH model.
We hope that our study will motivate further theoretical and experimental investigations on the KL with anisotropic exchange interactions in the future.

\begin{acknowledgments}
We thank H. Ueda, Y. Yamaji, T. Okubo, N. Kawashima, and M. Imada for useful discussions. 
This work was supported by MEXT, Japan, as a social and scientific priority issue (creation of new functional devices and high-performance materials to support next-generation industries) to be tackled by using a post-K computer. The numerical calculation was carried out at the facilities of the Supercomputer Center, the Institute for Solid State Physics, the University of Tokyo.
\end{acknowledgments}

%\bibliography{main}
\end{document}